# A simplified convolutional sparse filter for impulsive signature enhancement and its application to the prognostic of rotating machinery

Xiaodong Jia[1], Ming Zhao[1,2], Haoshu Cai, Jay Lee

[1] NSF I/UCR Center for Intelligent Maintenance Systems, Department of Mechanical Engineering, University of Cincinnati, PO Box 210072,Cincinnati, Ohio 45221-0072, USA

[2] School of Mechanical Engineering, Xi'an Jiaotong University, Xi'an 710049, China.

*Abstract*

Impulsive signature enhancement (ISE) is an important topic in the monitoring of rotating machinery and many different methods have been proposed. Even though, the topic of how to leverage these ISE techniques to improve the data quality in terms of prognostics and health management (PHM) still needs to be investigated. In this work, a systematic view for data quality enhancement is presented. The data quality issues for the prognostics and health management (PHM) of rotating machinery are identified, and the major steps to enhance data quality are organized. Based on this, a novel ISE algorithm is originally proposed, the importance of extracting scale invariant features are explained, and also related features are proposed for the PHM of rotating machinery. In order to demonstrate the effectiveness of the novelties, two experimental studies are presented. The final results indicate that the proposed method can be effectively employed to enhance the data quality for machine failure detection and diagnosis.

## 1. Introduction

Driven by the trend of industrial big data, Prognostics and Health Management (PHM) is now drawing more and more attention in both academia and industry(Lee et al., 2014a). Rotating machinery, which consists of a big family of industrial equipment, is one of the most important topics in PHM(Lee et al., 2014b). Different with Condition Based Monitoring (CBM), which mainly focuses on fault detection and diagnosis, prognosis is aiming at incipient machine failures and the prediction of Remaining Useful Life (RUL) (Lee et al., 2014b). Based on prognosis, health management develops optimal maintenance strategy for the upcoming failures. For rotating machines, effective PHM system should detect fault precursors, evaluate operation risks, predict the RUL of the equipment and autonomously optimize maintenance actions.

To ensure the success of PHM, good quality data from rotating machines is always critical (Chen, 2012). For PHM, data with sufficient quality should have good trend-ability, detect-ability and cluster-ability, which are posed by the needs of degradation assessment, incipient fault detection and prognosis respectively (Chen, 2012). However, this is challenging for rotating machinery. The reasons are manifold: 1) rotating machines are often working in harsh environment, the sensor readings from these machines often contain multiple sources of interferences and significant amount of noise; 2) the fault signatures of rotating machines often manifest as periodic impulses in the vibration, which is easily to be over-whelmed by unrelated harmonics; 3) Transmission path effect, which introduces additional sources of interferences, further diminishes the detect-ability of weak fault signatures (Jia et al., 2017b).

To address these challenges, different signal processing tools have been proposed for Impulsive Signature Enhancement (ISE) (Randall and Antoni, 2011, Zhao et al., 2013). Common approaches in frequency domain involve the Kurtogram (Antoni, 2007) and its variants(Xu et al., 2015), which is designed to find the most spiky frequency band using dyadic filter bank. Other ISE approaches as Empirical Mode Decomposition (EMD)(Lei et al., 2013) and Singular Value Decomposition (SVD)(Zhao and Jia, 2017) are employed to decompose the signal in first and then select the best related signal components. Wavelet analysis is a time-frequency domain approach which can highlight transients in the vibration series(Qiu et al., 2006). By reviewing these approaches, we would like to characterize two gaps in the perspective of PHM modelling. 1) Most of current ISE tools require a lot of trail-and-error efforts, which are challenging to be standardized for PHM analysis. 2) Most studies of ISE approaches stop after filtering out the fault signatures from vibration data. How to apply these tools for PHM analysis still needs more investigation.

To fill in these research gaps, this paper proposes a systematic methodology that can actively enhance the data quality for the PHM of the rotating machinery. The novelty of this paper includes the derivation of a simplified Convolutional Sparse Filter (CSF) for ISE and the design of scale invariant features for PHM modeling. The CSF, which is proposed in our previous work Ref.(Jia et al., 2017b), is a novel adaptive filter for vibration signal processing that requires few tuning efforts. This work simplifies the cost-function of the original CSF and boosts its efficiency. Scale invariant features are proposed later to extract reliable features from the CSF filtered signal for PHM. These features are proposed to evaluate the significance of periodic impulsive patterns in signal by sparse measures and period measures.

The rest of this paper is organized as follows. In section 2, the data quality issues for PHM are identified and discussed. In Section 3, the CSF from our previous work is revisited, and the simplified CSF is proposed. In Section 4, a methodology for data quality enhancement is proposed and the scale invariant features are discussed. In Section 5, two bearing case studies are presented. The conclusion remarks are given in Section 6.

## 2. Data quality issues in PHM

PHM is an engineering discipline that provides users with better transparency in regards to the health condition of machines and critical components. It serves as an agent that synchronizes factory shop floor and upper level management, and it essentially enables extended machine uptime and optimized shop floor management. In the real applications, the operation data from machines is not in always in perfect condition for analysis. Basically, the data quality issues in PHM can be categorized as:

(1) *Noise, missing value and incorrect records*: a. sensor data from machines usually carry ambient noise and unknown source of disturbances; b. missing values can happen during the data transmission and storage; c. incorrect records are common in the maintenance logs or archives.

(2) *Data insufficiency and redundancy:* it is difficult to judge whether the data from concerned system is adequate. One common way is to install numerous sensors at the initial stage. In this case, redundant information from different data channels needs to be removed.

(3) *Missing context information***:** one struggle of PHM is to isolate the machine deterioration from the ambient or working regime fluctuations. Therefore, context information is critical to identify similar operation conditions for PHM analysis. However, this context information is usually not archived or wrongly regarded as unimportant in real applications.

To better define desirable dataset for PHM analysis, we would like to present three additional desirable data features based on the needs of PHM analysis.

(1) **Good detect-ability** means the machine abnormalities can be well detected from normal conditions.

(2) **Good cluster-ability** means the observations from different failure modes have strong inclination to form clusters.

(3) **Good trend-ability** means the run-to-failure data collected at different stages of machine life indicates a strong link to the change of machine health condition.

For rotating machinery, data quality enhancement can be achieved by enhancing the impulsive signatures in vibration data and by extracting reliable features from the signal, which will be elaborated in later sections.

## 3. The simplified convolutional sparse filter

In this section, the CSF method for ISE will be revisited and the simplification of CSF will be introduced and illustrated via simulative examples.

### 3.1. Revisit of the convolutional sparse filter

The CSF is originally proposed in Ref. (Jia et al., 2017b) to recover the impulsive signature from noisy data. The idea of CSF is similar to the Minimum Entropy Deconvolution (MED) filter that has been studied in the monitoring of gears in recent years(Endo and Randall, 2007). The MED filter recovers the impulsive patterns in the vibration data by maximizing the Kurtosis. However, one disadvantage of using Kurtosis for sparse optimization is that it is too sensitive to the outliers. Therefore, the MED filter often quickly converges to the highest peak in the signal. To address this problem, the $l_1/l_2$ norm is proposed in Ref. (Jia et al., 2017b) to govern the sparse optimization and the proposed method is found more robust than MED.

In the formulation of CSF algorithm, the vibration signal $y \in \mathbb{R}^N$ is organized into a Hankel matrix as follows:

$$\mathbf{Y} = \begin{bmatrix} y_1 & y_2 & y_3 & \cdots & y_l \\ y_2 & y_3 & y_4 & \cdots & y_{l+1} \\ y_3 & y_4 & y_5 & \cdots & y_{l+2} \\ \vdots & \vdots & \vdots & \vdots & \vdots \\ y_{N-l+1} & y_{N-l+2} & y_{N-l+3} & \cdots & y_N \end{bmatrix} \tag{1}$$

The purpose of CSF is to find a feature mapping $\Phi: \mathbf{Y} \mapsto \mathbf{f} = \mathbf{Y} \cdot \mathbf{w}$, so that the $l_1/l_2$ norm of the feature matrix $\mathbf{f}$ can be minimized. In this manner, we expect to have sparse (or spiky) signal components to be enhanced. Graphically, this feature mapping process is equivalent to a neural network as shown in Fig. 1. The cost function of the neural network can be optimized using off-the-shelf L-BFGS packages. In the sense of explicitly, the cost function for the CSF is written as:

$$\min \sum_{i=1}^{M} \left\| \hat{\mathbf{f}}^{(i)} \right\|_{l_1} = \min \sum_{i=1}^{M} \left\| \frac{\overline{\mathbf{f}}^{(i)}}{\left\| \overline{\mathbf{f}}^{(i)} \right\|_{l_2}} \right\|_{l_1} \tag{2}$$

Where $\mathbf{f}$ denotes the feature matrix from the neural network and $M$ corresponds to the number of rows in the Hankel matrix. $\mathbf{f}_j^{(i)}$ in Eq.(2) represents the $j^{th}$ row (feature) for $i^{th}$ columns (sample) of the feature matrix and $\mathbf{f}_j^{(i)} = \left| w_j^T \cdot y^{(i)} \right|$. $\overline{\mathbf{f}}_j = \mathbf{f}_j / \left\| \mathbf{f}_j \right\|_{l_2}$ normalizes each feature by dividing the $l_2$ norm of each feature, and $\hat{\mathbf{f}}^{(i)} = \overline{\mathbf{f}}^{(i)} / \left\| \overline{\mathbf{f}}^{(i)} \right\|_{l_2}$ normalizes each sample by dividing the $l_2$ norm of each column in the feature matrix. Since the $l_1$ norm minimization in Eq. (2) is non-smooth and will introduce complexity in solving the minimization problem, a soft-absolute function $\mathbf{f}_j^{(i)} = \sqrt{\xi + \left( w_j^T \cdot y^{(i)} \right)^2} = \left| w_j^T \cdot y^{(i)} \right|$ is adopted to simplify the problem, where $\xi = 10^{-8}$.

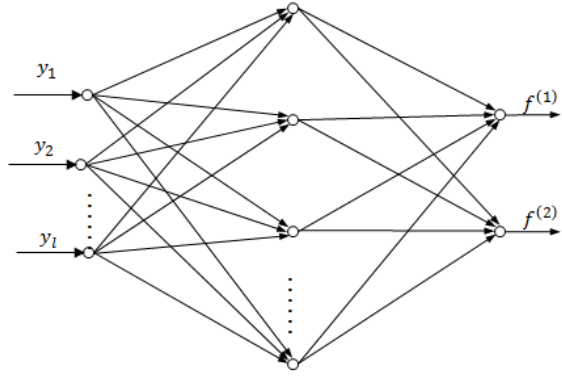

Fig. 1 The network structure of original CSF (Jia et al., 2017b)

**3.2. Simplified convolutional sparse filter**

Based on the previous discussion, it is easy to find that the original CSF is computationally expensive, especially when the multi-layer and multi-node structure is adopted. To simplify the CSF, its cost-function is re-formulated in this paper. The cost-function in Eq.(2) is originally proposed in (Ngiam et al., 2011) to extract sparse features from natural images. Therefore, it includes a feature-wised normalization for lifetime sparsity and a sample-wised normalization for population sparsity. In the application of ISE, we find the sample-wised normalization unnecessary. This is because the exact position of one or several impulses cannot give much useful diagnostic information, only the relative period between impulses is indicative to certain machine failures.

The simplification of original CSF is achieved by removing the constraints of population sparsity in Eq.(2), which can be written as:

$$\min \sum_{i=1}^{M} \left\| \bar{\mathbf{f}}^{(i)} \right\|_{l_1} = \min \sum_{i=1}^{M} \frac{\left\| \mathbf{f}_j \right\|_{l_1}}{\left\| \mathbf{f}_j \right\|_{l_2}} \qquad (3)$$

Numerical test indicates that Eq.(3) can be further simplified by setting the number of features $M = 1$. Therefore, the optimization problem for the simplified CSF can be re-formulated as:

$$\min \ J_{1,2}(\mathbf{f}) = \min_{\mathbf{f}} \frac{\|\mathbf{f}\|_{l_1}}{\|\mathbf{f}\|_{l_2}} = \min_{\mathbf{f}} \frac{\sum_{i=1}^{N}|f_i|}{\left(\sum_{i=1}^{N}|f_i|^2\right)^{1/2}} \quad (4)$$

$$\text{s.t.} \quad \mathbf{f} = \mathbf{Y} \cdot \mathbf{w}, \ \|\mathbf{w}\|_{l_2} = 1, \ |f_i| = c_i$$

In Eq.(4), the output $\mathbf{f}$ is a feature vector, $f_i$ denotes the $i$-th element in the feature vector and the absolute of $f_i$ is estimated through the soft-absolute function mentioned earlier. Optimization of Eq.(4) can be achieved by the off-the-shelf L-BFGS packages, and the first order derivative of the cost-function can be stated as:

$$\frac{\partial J_{1,2}}{\partial w_j} = \sum_i \frac{\partial J_{1,2}}{\partial c_i} \cdot \frac{\partial c_i}{\partial f_i} \cdot \frac{\partial f_i}{\partial w_j} = \sum_i \left\{ \left[ \frac{1}{\sum_i (c_i^2)^{\frac{1}{2}}} - \frac{\left(\sum_i c_i\right) \cdot c_i}{\left(\sum_i c_i^2\right)^{\frac{1}{2}+1}} \right] \cdot \frac{f_i}{\sqrt{f_i^2 + \varepsilon}} \cdot x_{i+j-1} \right\} \quad (5)$$

As a summary, the simplifications made to the original CSF can be better read from Table. 1. In the simplified CSF, the previous multiple-layer and multi-node neural network structure is replaced with a signal neuron. In the step 3 in Table. 1, the population sparsity constraint in Eq.(2) is removed in Eq.(4). Therefore, the over-all optimization process is significantly simplified. In addition, the step 4 in the original CSF is skipped, since the output of step 3 in the simplified CSF is already a real-value signal instead of a matrix.

Table. 1 Algorithm table for the original CSF and the simplified CSF

| **Algorithm 1: the original CSF (Jia et al., 2017b)** | **Algorithm 2: The simplified CSF** |
|---|---|
| *Step 1*: formatting input signal y into Hankel matrix **Y** | *Step 1*: formatting input signal *y* into Hankel matrix **Y** |
| *Step 2*: Initializing a **multi-layer** and **multi-node** neural network for ISE | *Step 2*: Initializing a **single-layer** and **single-node neuron** for ISE |
| *Step 3*: Optimizing the **neural network** by minimizing the cost-function in **Eq.(2)** | *Step 3*: Optimizing the **neuron** by minimizing the cost-function in **Eq.(4)** |
| *Step 4*: Preserving the dominant features in the output of step 3 by using **dimension reduction** techniques, like PCA or diffusion map. | |

### 3.3. Simulative examples

In order to test the performance of the simplified CSF, three simulative studies are performed in this section.

*a. Outlier sensitivity test*

In the first simulative study in Fig. 2, the input signal in Fig. 2(a) is simulated as random Gaussian noise. Similar to the result given by the original CSF in Ref. (Jia et al., 2017b), the output of the simplified CSF does not converge to the outlier. In comparison, the MED filter converges to the highest peak in the input signal. This example demonstrates that the proposed method is more robust than MED filter.

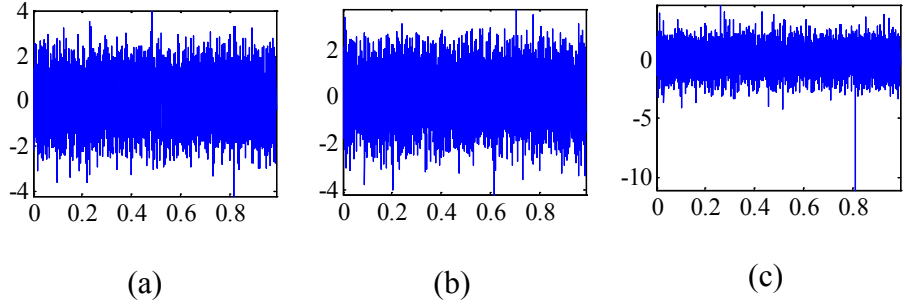

(a) (b) (c)

Fig. 2 Simulative case study 1. (a) Gaussian noise; (b) the filtering result given by simplified CSF; (d) the filtering result given by MED.

### b. *Simulative study for bearing out-race fault and inner-race fault*

Fig. 3 and Fig. 4 demonstrate another two case studies to further compare the performance of the original CSF and the simplified CSF. The input signal in Fig. 3 (a) simulates a bearing outer race fault. Fig. 3(c) and Fig. 3(e) plot the filtered vibration signal given by the original CSF and the simplified CSF. Both of which successfully enhanced the impulsive pattern in the input signal. Further comparison in the envelop spectrum in Fig. 3(b), (d) and (f) indicates that the bearing fault signature is correctly enhanced. Similarly, the case study in Fig. 4, which simulates the vibration signal for bearing inner race fault, also supports the conclusion that the simplified CSF in this study can achieve comparable results in regards to the fault signature enhancement. In both case studies, the signal-to-noise ratio (SNR) in the simulated signal is -8dB. For other details about how to simulate signals, interest readers can refer to Ref. (Jia et al., 2017b).

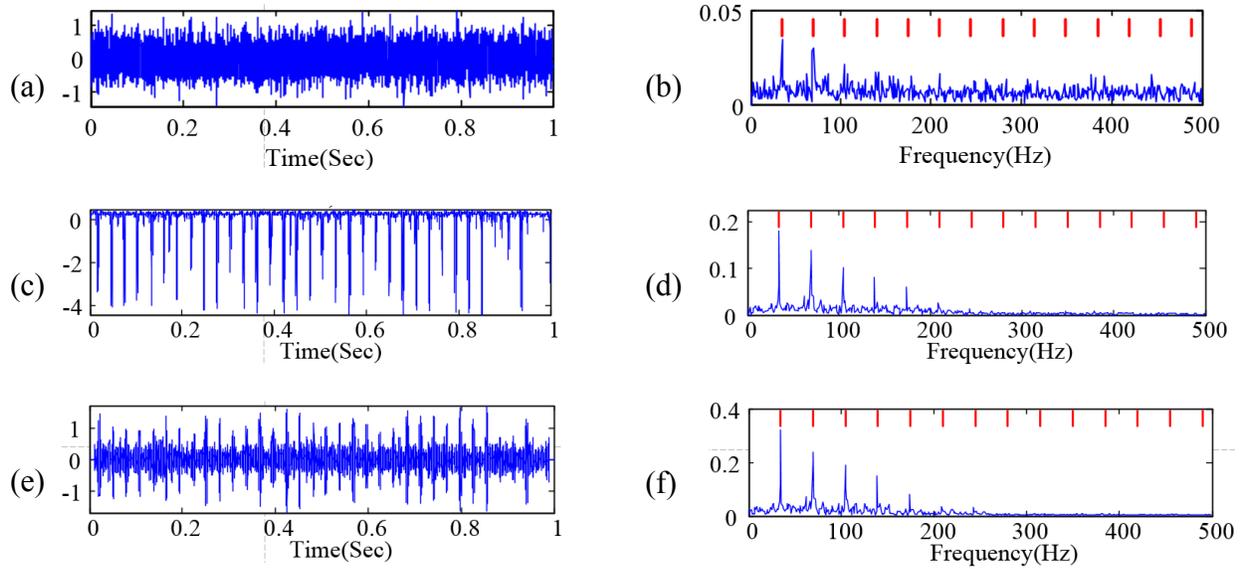

Fig. 3 Simulative case study 2: (a) simulated vibration signal for bearing outer-race fault; (b) traditional envelop spectrum of the simulated data; (c) time waveform of the principal feature from original CSF; (d) frequency spectrum of the principal feature from original CSF; (e) the filtered signal from simplified CSF; (f) envelop spectrum of the filtered signal from simplified CSF; Red lines in (b), (d), (f) represent the fault harmonics.

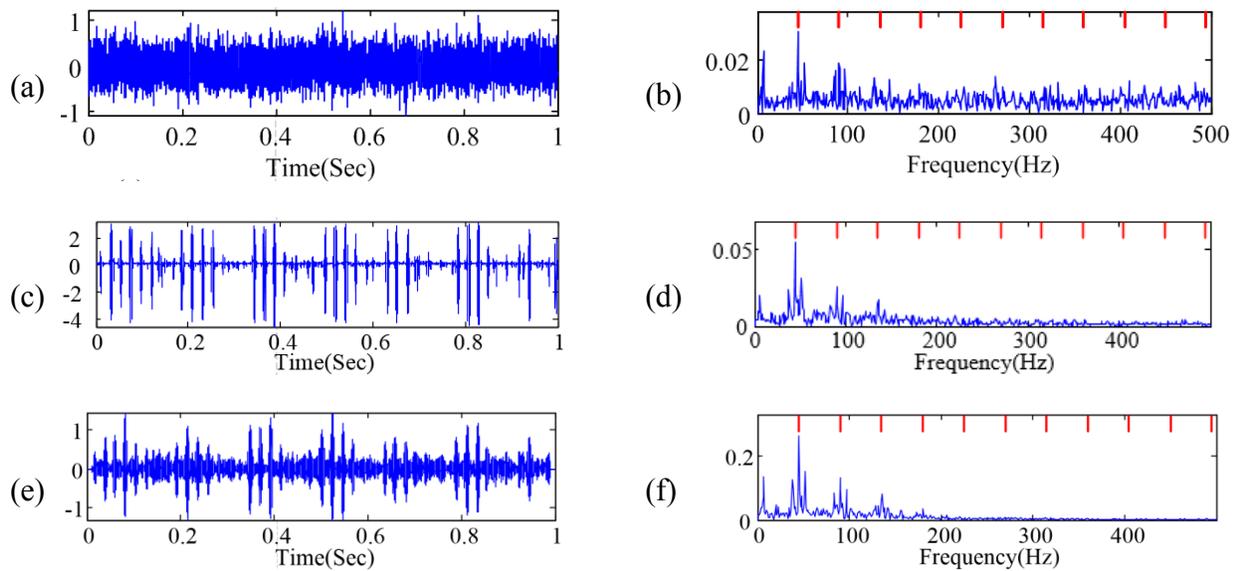

Fig. 4 Simulative case study 3: (a) simulated vibration time series for bearing inner race fault; (b) traditional envelop spectrum of the simulated data; (c) time waveform of the first principal

feature from original CSF; (d) frequency spectrum of the first principal feature from original CSF; (e) the filter signal from the simplified CSF; (f) envelop spectrum of the filtered signal from the simplified CSF; Red lines in (b), (d), (f) represent the fault harmonics.

*c. Computational efficiency test*

The efficiency of the simplified CSF is benchmarked against the original CSF in Table. 2. Values in Table. 2 are obtained by averaging the computation time of 20 runs using the same input signal. These results indicate that the computation time is significantly improved by the simplified CSF. In all the simulations, a 2-layer CSF is employed and the number of nodes for each layer is set to 150, the structure of the network is the same with that in Ref. (Jia et al., 2017b).

Table. 2 Comparison of the computation efficiency of the original CSF and the proposed CSF

|  | The original CSF(Jia et al., 2017b) (Sec) | The simplified CSF (Sec) |
|---|---|---|
| Simulative case study 1 | 227.6402 | 0.5028 |
| Simulative case study 2 | 236.5595 | 0.5028 |

## 4. Methodology

In this section, the Scale Ambiguity Phenomenon (SAP) in feature extraction is firstly discussed and corresponding solutions are suggested. Based on this, a systematic methodology for data quality enhancement is proposed, its connections with the novelties of this work is stated.

### 4.1. Scale invariant features

Scale invariant features are proposed to fix the SAP that exists in the filtered signal. The SAP is introduced by the convolutional filtering process, and it can be stated as that the scale of the oscillation pattern in the input series become ambiguous after convoluting with the filter coefficients. To understand SAP, one can find that the filtered signal in Fig. 3(c) and (e) are different with the vibration scale of the input series in Fig. 3(e). This is because that the convolutional filtering process can only highlight the impulsive pattern in the input signal. It is almost impossible to exactly recover the scale of impulsive pattern without addition information.

Scale invariant features provide a simple way to solve this undesirable phenomenon. Scale invariant features are intrinsically scale irrelevant. If we define the feature extraction process as mapping $\mathbb{R}^N \mapsto \mathbb{R}$, then the scale invariant features can be described as:

$$\Psi(\mathbf{f}) = \Psi(k \cdot \mathbf{f}), \quad k \neq 0 \tag{6}$$

Where $\mathbf{f} \in \mathbb{R}^N$ denotes the filtered signal, $k$ represents arbitrary scale factor except 0. Eq. (6) basically states that the value of scale invariant feature is not affected by the scale change of the filtered signal.

### 4.2. Suggested scale invariant features

In this study, we would like to suggest several scale invariant features that are useful to the rotating machinery. One important family of scale invariant features involve the generalized $l_p/l_q$ norm that is employed to evaluate the sparseness/spikiness of the signal (Jia et al., 2017a). The generalized $l_p/l_q$ norm can be inferred as scale invariant since as follows

$$J_{p,q}(k \cdot \mathbf{f}) = \left( \frac{\|k \cdot \mathbf{f}\|_{l_p}}{\|k \cdot \mathbf{f}\|_{l_q}} \right)^p = \left( \frac{k \|\mathbf{f}\|_{l_p}}{k \|\mathbf{f}\|_{l_q}} \right)^p = \left( \frac{\|\mathbf{f}\|_{l_p}}{\|\mathbf{f}\|_{l_q}} \right)^p = J_{p,q}(\mathbf{f}) \tag{7}$$

It is also noticed that the $l_1/l_2$ norm that is discussed in the simplified CSF is just a special case of generalized $l_p/l_q$ norm when $p = 1, q = 2$. Similarly, Kurtosis also belongs to the family of $l_p/l_q$ norm if we set $p = 2, q = 4$

Envelop harmonic-to-noise ratio (EHNR) is another type of scale invariant feature for periodicity evaluation. EHNR is defined as the auto-correlation function (ACF) of the signal envelop, see Ref. (Xu et al., 2016). Therefore, it is easy to find that the EHNR is scale invariant since the ACF in the EHNR is scale invariant. Knowing that EHNR is a scale invariant criterion to evaluate the periodicity of signal envelop, we would like to further customize this criterion into Band-Limited EHNR (BLENHR). The BLEHNR is designed to search the highest value in the ACF of signal envelop within the specified time band. Taking rolling element bearing for example, Ball Passing Frequency at Outer-race (BPFO), Balling Passing Frequency at Inner-race (BPFI) and Ball Spin Frequency (BSF) are three typical fault frequencies that correspond to three different types of failure. Then the BLEHNR features can be extracted as the highest values within the specified time bands near 1/BPFO, 1/BPFI and 1/BSF;

### 4.3. Methodology for data quality enhancement

The proposed methodology for PHM data quality enhancement for rotating machinery can be described as three major steps. (1) Impulsive pattern enhancement: the purpose of this step is to enhance the repetitive impacts introduced by mechanical defects. (2) Scale invariant feature extraction: the purpose of extracting scale invariant features is to eliminate the scale invariant phenomenon. For rotating machinery, the effective feature should be able to evaluate the spikiness and the periodicity of the signal without the influence of signal scales. (3) PHM data modeling: in this step, different modeling techniques for health diagnosis, health assessment and health prediction will be employed to obtain useful health information from the features.

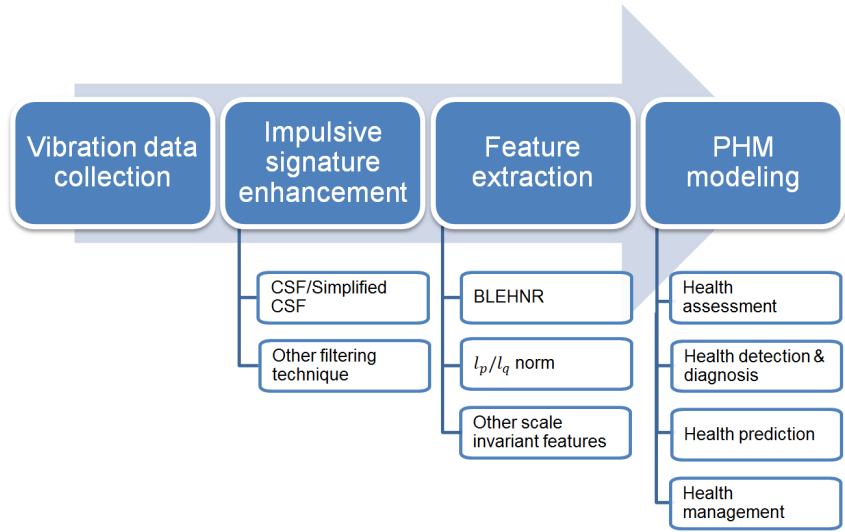

Fig. 5 The vibration based data quality enhancement methodology for rotating machinery

A more general flow of the proposed methodology is in Fig. 5. Within each functional module in Fig. 5, the simplified CSF and the scale invariant features that are discussed previously serves as candidate algorithms. It is worth mentioning that the candidate algorithms in Fig. 5 are obviously not exhaustive, interest readers can further add other algorithms, like MED filter, Kurtogram, etc.

## 5. Results and discussion

In this section, two experimental case studies are investigated. In case study 1, a run-to-failure dataset provided by NASA website (Qiu et al., 2006) is employed to show the capability of the proposed method for health assessment. In case study 2, a dataset with multiple bearing failure modes are employed to show the strength of the proposed method for health diagnosis.

### 5.1. Experimental case study 1: incipient fault detection for rolling element bearing

In this run-to-failure experiment, four Rexnord ZA-2115 double row bearings are studied, and a radial load of 6000 lbs is added to the bearing by a spring mechanism. The rotating speed of the shaft was kept at 2000 rpm throughout the experiment and the data was sampled at 20kHz. The vibration is collected every 20 minutes by a National Instruments DAQCard-6062E data acquisition card and the data duration for each test file is 20480 points. At the late stage of bearing life, an outer-race bearing fault was discovered and 984 test files are collected in total in this experiment.

In this experiment, an incipient bearing failure is detected as early as in test file No.540 (Xu et al., 2016). The time domain vibration series in Fig. 6(a) and Fig. 6(c) present the test file No.540 and No.600 respectively, from which it is hard to visualize any repetitive pattern directly. After applying the simplified CSF, the repetitive pattern that has been masked by ambient noise is successfully enhanced. The results in Fig. 6 further support the effectiveness of the simplified CSF for impulsive signature enhancement.

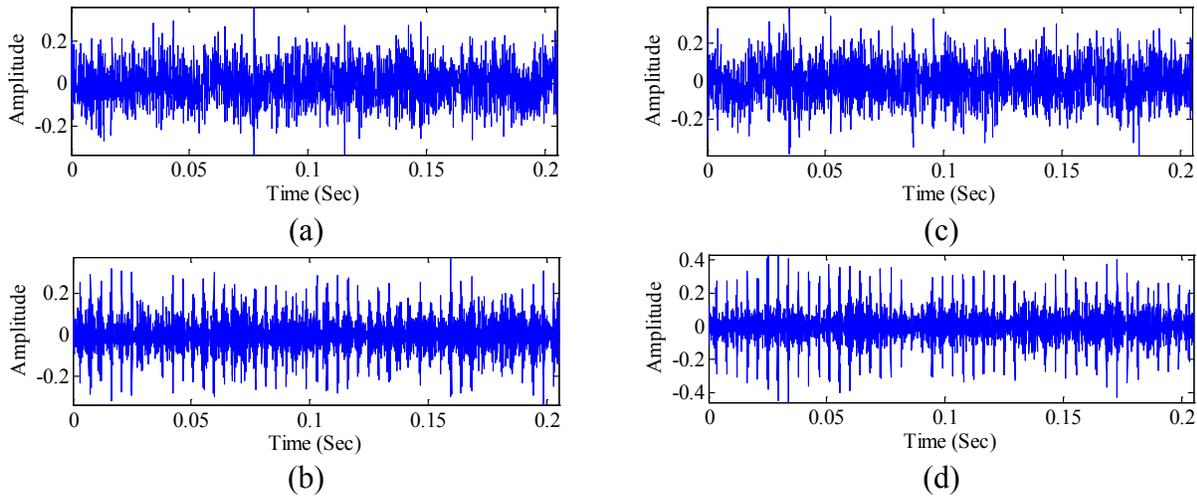

Fig. 6 The original vibration series at test file No.540 (a) and test file No.600(c). The detection result given by the simplified CSF at test file No. 540 (b) and test file No.600.

To further demonstrate the data quality enhancement made by the simplified CSF, scale invariant features signal are extracted and compared, and a health assessment model for the bearing is built using self-organizing map (SOM) – minimum quantization error (MQE). Fig. 7~ Fig. 9 presents the comparison of several scale invariant features extracted from the scenarios with/without applying the simplified CSF. These results clearly indicate that the incipient failure of the bearing can be detected with higher confidence when the simplified CSF is applied. After feature extraction, all these features are formatted into a feature matrix, and the first 20 test samples are employed to train the SOM model. In the testing process, the deviation between each testing sample and the training samples are evaluated using the MQE value, which is plotted in Fig. 10. Comparison of the MQE values in Fig. 10 indicates that the bearing failure can be better detected and the incipient bearing faults can be detected with enhanced confidence. Therefore, the detect-ability of the bearing abnormalities is significantly enhanced.

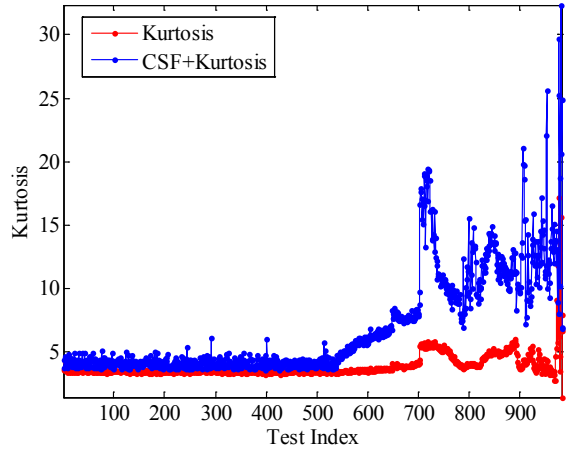
Fig. 7 Comparison of Kurtosis

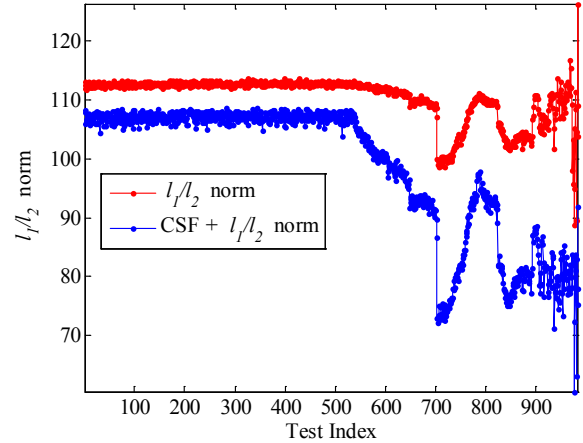
Fig. 8 Comparison of $l_1/l_2$ norm

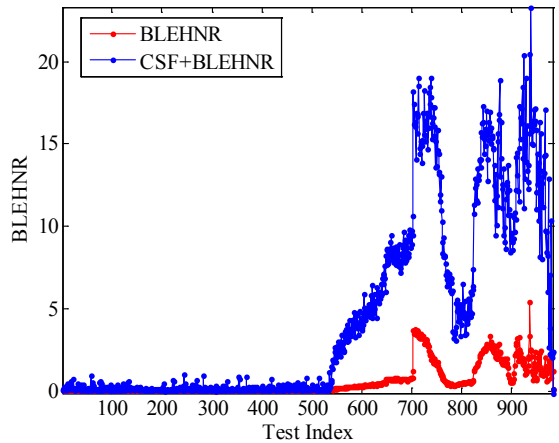
Fig. 9 Comparison of BLEHNR

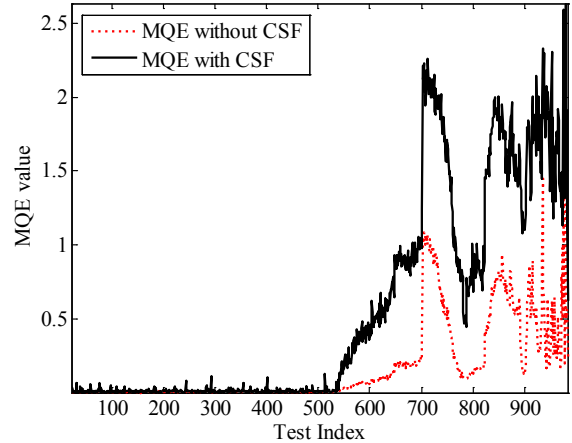
Fig. 10 Comparison of BLEHNR

## 5.2. Experimental case study 2: multiple fault classification for rolling element bearing

To demonstrate the data quality improvements in terms of multiple fault classification, a dataset with multiple bearing failure modes are employed. This dataset is collected from a test-bed at center for Intelligent Maintenance System (IMS) lab(Chen et al., 2013). In this experiment, the spindle rotation speed is 800rpm and 8 different bearing failures are listed in Table. 3.

Table. 3 the bearing failure conditions

| Failure mode | Condition description |
|---|---|
| F1 | Normal |
| F2 | Outer-race fault |
| F3 | Inner-race fault |
| F4 | Roller fault |
| F5 | Inner-race + roller fault |
| F6 | Outer-race + inner-race fault |
| F7 | Outer-race + roller fault |
| F8 | Outer-race + inner-race + roller fault |

In this study, two feature matrix with 5 different features, as in Table. 4, are extracted from the raw data and the CSF filtered data. The first two features in Table. 4 are used to evaluate the spikiness of the signal and the rest three features are used to evaluate the periodicity at corresponding fault frequencies.

Fig. 11(a) visualizes the clustering tendency of the raw data in the principal component (PC) space. The principal features in Fig. 11(a) are extracted by PCA from raw features. Even though Fig. 11(a) clearly indicates 8 different clusters in the PC space, several outliers can be still spotted and circled out. And also, the cluster F1 and cluster F6 are distributed closely with each other. In comparison, the filtered data in the PC space in Fig. 11(c) indicates that all the clusters are better separated and no outliers are visualized. Further clustering analysis on the PC values in Fig. 11(c) using k-means indicates no mis-clustering issues. Therefore, the cluster-ability of the dataset is improved.

Table. 4 Selected features for bearing fault diagnosis

| # | Selected Features |
|---|---|
| 1 | Kurtosis |
| 2 | $l_1/l_2$ norm |
| 3 | BLEHNR at $1 \times$ BPFO |
| 4 | BLEHNR at $1 \times$ BPFI |
| 5 | BLEHNR at $1 \times$ BSF |

Fig. 11(b) and Fig. 11(d) present the results from Visual Assessment of clustering Tendency (VAT) approach. VAT is proposed in Ref. (Bezdek and Hathaway, 2002, Wang et al., 2010). It visualizes the clustering tendency of the dataset by reordering the dissimilarity matrix generated from the feature matrix. Implementation details about VAT approach can be found in Ref.(Chen et al., 2013). Fig. 11(b) shows the VAT image from the raw data. Each dark block in diagonal position in Fig. 11 represents that these data samples are closely distributed and can be regarded as one cluster. In Fig. 11(b), most of the data samples are correctly clustered except that two of them are wrongly regarded as outliers. This observation matches with the PCA analysis in Fig. 11(a). In comparison, VAT image in Fig. 11(d) for the filtered signal clusters all the data samples correctly. The results in Fig. 11 clearly indicate that the data quality of the original vibration data is successfully enhanced.

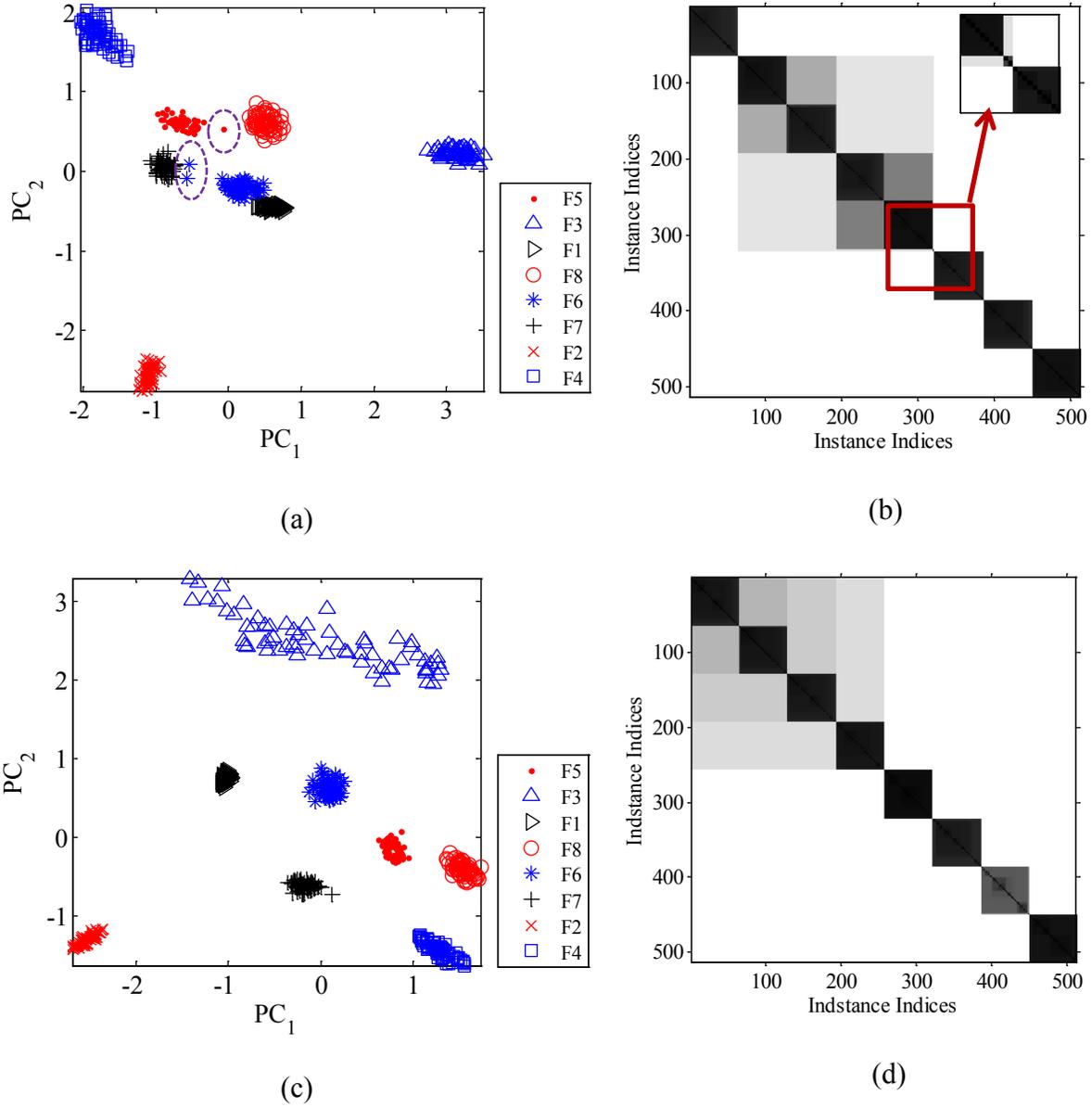

Fig. 11 Results for data quality enhancement for multiple failure classification. (a) The scatter plot of principal features from raw data. (b) The VAT image of the extracted features from the raw data. (c) The scatter plot of principal feature from the filtered data by simplified CSF. (d) The VAT image VAT image of the extracted features from the filtered data by simplified CSF.

## 6. Conclusion

In this study, a systematic methodology for data quality enhancement is proposed. The proposed methodology is targeting at enhancing the vibration data quality for the PHM of rotating machinery. Within this methodology, the simplified CSF is proposed and the scale invariant features are investigated. In addition, two families of scale invariant features, the generalized $l_p/l_q$ norm and the Band-Limited EHNR, are suggested. The effectiveness of the novelties in this work is justified through two experimental studies, one for the health assessment of rolling element bearing and the other for the multiple fault classification of bearing. In this study, several conclusions can be achieved. (1) The simplified CSF can be effectively employed for impulsive pattern enhancement. (2) The BLEHNR and $l_p/l_q$ norm are effective features for the health analysis of rotating machinery. (3) The results from the experimental studies clearly indicate that the proposed methodology can be promising for data quality enhancement.

In future works, application of the proposed methodology in gears and other unknown engineering system will be interested. And also, enrichment of the candidate algorithms within the framework in Fig. 5 will be further investigated.

**Acknowledgement**

This research is supported by the National Natural Science Foundation of China (Grant No. 51405373), and the China Postdoctoral Science Foundation (2014M562400), which are highly appreciated by the authors.